\documentclass[a4paper,10pt]{article}

\usepackage{amsmath}
\usepackage{amsthm}
\usepackage{amssymb}
\usepackage{eepic}

\newlength{\fuyasu}
\theoremstyle{definition}
\newtheorem{example}{Example}

\newcommand{\gehn}{\ensuremath{\mathfrak{g}_n}}

\newcommand{\ot}{\otimes}
\newcommand{\ol}[1]{\overline{#1}}
\newcommand{\Z}{{\mathbb Z}}

\newlength{\BallWidth} 
\newcommand{\ballmovingarrow}[1]{
\path(0,0)(0,-0.5)(#1,-0.5)(#1,-1)
\put(#1,-1){\vector(0,-1){0}}}

\newcommand{\yajirusiStep}[1]{
\qbezier(0.2,0)(1.7,-1)(0.2,-2)
\put(0.2,-2){\vector(-3,-2){0}}
\put(1.2,-1){\makebox(0,0)[l]{Step #1}}}

\newcommand{\batu}[1]{
{\setlength{\unitlength}{10pt}
	\put(-1,0.3){\vector(1,0){2.4}}
	\put(0.6,0.3){\line(-1,1){0.6}}
	\put(0.375,0.75){\makebox(0,0)[lb]{{\tiny$#1$}}}
	{\thicklines\put(0,1.5){\vector(0,-1){2.5}}}
}}

\newcommand{\yajirusiK}[1]{
$K_{#1}$\;
{\setlength{\unitlength}{10pt}
	{
	\qbezier(0.5,1.5)(0,1)(0,0)
	\qbezier(0,0)(0,-0.5)(0.5,-1)
	\put(0.5,-1){\vector(3,-4){0}}
	}
}}

\newcommand{\batten}[5]{%
\begin{picture}(40,40)(-20,-20)
	\put(-10,0){\vector(1,0){20}}
	\put(0,4){\line(1,-1){4}}
	\put(2.5,2.5){\makebox(0,0)[lb]{{\tiny$#5$}}}
	\thicklines
	\put(0,10){\vector(0,-1){20}}
	\put(-11,0){\makebox(0,0)[r]{$#1$}}
	\put(0,11){\makebox(0,0)[b]{$#2$}}
	\put(0,-11){\makebox(0,0)[t]{$#3$}}
	\put(11,0){\makebox(0,0)[l]{$#4$}}
\end{picture}
}

\title{Simple Algorithm for \\
Factorized Dynamics of $\gehn$-Automaton}

\author{
Goro Hatayama\thanks{
Institute of Physics, University of Tokyo, Komaba, Tokyo 153-8902, Japan},
Atsuo Kuniba,$\hspace{-1.2mm}^*$
and Taichiro Takagi\thanks{
Department of Applied Physics, National Defense Academy,
Yokosuka 239-8686, Japan}
}

\date{}
\begin{document}
\maketitle

\begin{abstract}
We present an elementary algorithm for the dynamics
of recently introduced soliton cellular automata associated with 
quantum affine algebra $U_q(\gehn)$ at $q=0$.
For $\gehn = A^{(1)}_n$, the rule reproduces  
the ball-moving algorithm in Takahashi-Satsuma's box-ball system.
For non-exceptional $\gehn$ other than $A^{(1)}_n$, 
it is described as a motion of particles and anti-particles 
which undergo pair-annihilation and creation through a neutral bound state.
The algorithm is formulated without using   
representation theory nor crystal basis theory.
\end{abstract}

\section{Introduction}\label{sec:1}

The box-ball system invented by Takahashi and Satsuma \cite{TS,T}
is a remarkable example of one dimensional soliton cellular automata.
It represents the dynamics of balls hopping along the array of 
boxes under a certain rule.
The evolution equations have been studied extensively in 
\cite{TTMS,TTM,TNS} by means of the ultradiscretization of 
soliton equations.

In \cite{HKT1,FOY,HHIKTT}, the box-ball system was identified 
with a solvable vertex model in statistical mechanics \cite{B} in a 
low temperature limit.
Mathematically it brought a crystal basis theory \cite{K}
into the game, and led to a generalization labeled with 
affine Lie algebras $\gehn$.
We call the resulting system the $\gehn$-automaton.
See \cite{HKT1} for the construction for non-exceptional $\gehn$, 
and \cite{HKOTY} for the study of soliton scattering.
The box-ball system corresponds to the $\gehn = A^{(1)}_n$ case \cite{FOY,HHIKTT}.

The dynamics of the $\gehn$-automaton 
is governed by the combinatorial $R$ \cite{KMN} in the crystal theory, 
which specifies the local interaction of a box and a carrier
in the language of the box-ball system.
Except the $A^{(1)}_n$ case \cite{NY},
the complexity of the combinatorial $R$ \cite{HKOT1,HKOT2} 
makes it tedious to compute the dynamics of the associated $\gehn$-automaton.
Fortunately the difficulty was overcome in \cite{HKT2} for the automaton having a  
carrier with infinite capacity, where a factorization 
of the time evolution into simple Weyl group reflections
was proved.

The aim of this paper is to translate the factorized dynamics  \cite{HKT2} 
stated in the crystal language into an analogue of the ball-moving algorithm,
and thereby to describe the $\gehn$-automaton in terms of a simple evolution rule.
For simplicity we shall restrict ourselves to the case 
where all the boxes  have minimal capacity.
In the crystal formulation of the time evolution: 
$B_M \ot (\cdots \ot B_{m_i}\ot B_{m_{i+1}} \ot \cdots ) 
\simeq (\cdots \ot B_{m_i}\ot B_{m_{i+1}} \ot \cdots ) \ot B_M$,
the situation corresponds to the choice $\forall m_i = 1$ and $M = \infty$.
The factorized dynamics for $A^{(1)}_n$-automaton reproduces the decomposition of 
the ball-moving algorithm in the box-ball system into a finer procedure  
where one only touches the balls with a fixed color.
For the other $\gehn$'s in question, 
the dynamics is decomposed similarly, where each finer procedure 
(denoted by $K_j$ in Section \ref{sec:2}) is now described as  
the motion of particles and anti-particles with a fixed color 
exhibiting pair-annihilation and creation through a neutral bound state.

In Section \ref{sec:2} we specify the set of local states and 
give the algorithm.
In Section \ref{sec:3} we present examples.
In Section \ref{sec:4} we list the patterns that behave as solitons.
In Section \ref{sec:5} we briefly explain the 
relation of the algorithm in Section \ref{sec:2} and the result in \cite{HKT2}.
The key is the formula (\ref{eq:relation}).

Although our notations originate in the crystal theory,
the contents in Sections \ref{sec:2}-\ref{sec:3} use no result 
{}from it.
Despite being just a translation of a more general result in \cite{HKT2},
we hope this paper possesses its own role to make 
the $\gehn$-automaton more accessible and familiar to the people 
working on cellular automata and discrete integrable systems.

\section{Algorithm}\label{sec:2}

We formulate the $\gehn$-automaton associated with 
the non-exceptional affine Lie algebra 
$\gehn = A^{(1)}_n (n \ge 1), A^{(2)}_{2n-1} (n \ge 3),
A^{(2)}_{2n} (n \ge 2), B^{(1)}_n (n \ge 3), C^{(1)}_n (n \ge 2),
D^{(1)}_n (n \ge 4)$ and $D^{(2)}_{n+1} (n \ge 2)$.
Actually the algorithm itself given below is meaningful 
formally for $A^{(2)}_3, B^{(1)}_2, D^{(1)}_2$ and $D^{(1)}_3$ as well.
First we specify the set $B$ of local states 
and the sequence $j_1, \ldots, j_d \in B$ as follows.

\begin{center}
\begin{tabular}{c|c|c}
\gehn & $B$ & $(j_d,\dots,j_1)$ \\\hline 
$A^{(1)}_{n}$
&$\left\{1,2, \dots, n+1  \right\}$
&$(2,3,\dots,n+1)$\\[5pt]
$A^{(2)}_{2n-1}$
&$\left\{ 1,2,\dots,n,-n,\dots,-2,-1  \right\}$
&$(2,3,\dots,n,-1,-n,\dots,-3,-2)$\\[5pt]
$A^{(2)}_{2n}$
&$\left\{ 1,2,\dots,n,-n,\dots,-2,-1,\emptyset  \right\}$
&$(2,3, \dots ,n, -1,-n, \dots ,-3, -2, \emptyset)$\\[5pt]
$B^{(1)}_n$
&$\left\{ 1,2,\dots,n,0,-n,\dots,-2,-1  \right\}$
&$(2, 3, \dots, n, 0, -n, \dots, -3, -2)$ \\[5pt]
$C^{(1)}_n$
&$\left\{ 1,2,\dots,n,-n,\dots,-2,-1  \right\}$
&$(2, 3, \dots, n, -1, -n, \dots, -3, -2, -1)$ \\[5pt]
$D^{(1)}_n$
&$\left\{ 1,2,\dots,n,-n,\dots,-2,-1  \right\}$
&$(2, 3, \dots, n, -n, \dots, -3, -2)$ \\[5pt]
$D^{(2)}_{n+1}$
&$\left\{ 1,2,\dots,n,0,-n,\dots,-2,-1,\emptyset  \right\}$
&$(2, 3, \dots, n, 0, -n, \dots, -3, -2, \emptyset)$ \\[5pt]
\end{tabular}
\end{center}
\vskip1em
Here $\emptyset$ should be understood as a symbol and not an empty set.
By the definition $d=n$ for $A^{(1)}_n$, 
$d=2n-2$ for $D^{(1)}_n$, $d=2n-1$ for $A^{(2)}_{2n-1}, B^{(1)}_n$, 
$d=2n$ for $A^{(2)}_{2n}, C^{(1)}_n$ and $D^{(2)}_{n+1}$.

We let 
\begin{equation}\label{eq:W}
W = \{ (\ldots, b_i, b_{i+1},\ldots) \in 
\cdots \times B \times B \times \cdots \mid b_i = 1 
\text{ for } \vert i \vert \gg 1 \}
\end{equation}
be the set of states of the automaton at a fixed time.
Setting 
\begin{equation*} 
J = \{ j_1, \ldots, j_d \},
\end{equation*}
one finds 
\begin{equation*} 
J = \begin{cases} B\setminus \{1,-1 \} & \text{ if } 
\gehn = B^{(1)}_n, D^{(1)}_n, D^{(2)}_{n+1},\\
B\setminus \{1 \} & \text{ otherwise }.
\end{cases}
\end{equation*}
For $j \in J$ (hence $j \neq 1$), we define 
\begin{equation}\label{eq:bar}
j^\ast = \begin{cases}
j & \text{ if } j \in \{ 0, -1, \emptyset\}, \\
-j & \text{ if } j \in J \setminus \{0,-1, \emptyset \}.
\end{cases}
\end{equation}
The time evolution of the $\gehn$-automaton 
$T: W \rightarrow W$ takes the factorized form 
\begin{equation}\label{eq:T}
T = K_{j_d} \cdots K_{j_2} K_{j_1}.
\end{equation}
To define the operator $K_j: W \rightarrow W\; (j \in J)$, 
one regards an element $(\ldots, b_i, b_{i+1}, \ldots)$ of $W$ 
as the one dimensional array of boxes containing $b_i$ in the $i$-th box.
(The $(i+1)$-th box to the right of the $i$-th box.)
In what follows we regard the box containing $1 \in B$ as 
an empty box.
Thus, taking some $b \, (\neq 1) \in B$ away {}from a box means the change 
of the local states $b \rightarrow 1$.
Similarly, putting $b \, (\neq 1) \in B$ into an empty box means 
the change $1 \rightarrow b$.
Note that the boundary condition in (\ref{eq:W}) says that 
only finitely many boxes are non-empty.
Under this convention the operator $K_j$ is defined by the following algorithm
which consists of 4 Steps:
\begin{enumerate}
\item Replace each $-1$ by a pair $j, j^\ast$ within a box.
\item Pick the leftmost $j$ (if any) and move it to the nearest right 
box which is empty or containing just $j^\ast$.
(Boxes involving the pair $j,j^\ast$ are prohibited as the destination.)

\item  Repeat Step 2 for those $j$'s that are not yet moved
    until all of $j$'s are moved once.
\item Replace the pair $j, j^\ast$ within a box (if any) by $-1$.
\end{enumerate}

\vskip0.2cm
\noindent
{\em Remark 1}.
For $\gehn = A^{(1)}_n$ where no $-1$ is present in $B$ and $J$, 
Step 1 and 4 become void and the above procedure shrinks to the well known 
ball-moving algorithm in the box-ball system \cite{T}.
\vskip0.15cm
\noindent 
{\em Remark 2}. When $j \in \{-1,0,\emptyset \}$, one can have 
two $j$'s within a box due to (\ref{eq:bar}).
As for those duplicated $j$'s,  
it does not matter which is left or right, but of course 
they should be distinguished according to whether they are moved or not yet moved.
See $K_{\emptyset}$ in Example \ref{ex:A24}.

\vskip0.15cm
\noindent
{\em Remark 3}. A little inspection shows that $K_{-1}$ actually admits a 
simpler description which is essentially the ball-moving algorithm
in the box-ball system:

\noindent \hskip0.5cm 1'. Pick the leftmost $-1$ (if any) 
and move it to the nearest right empty box.

\noindent \hskip0.5cm 2'. 
Apply Step 1' for those  $-1$'s  that are not yet moved (if any).

\noindent \hskip0.5cm 3'. Repeat Step 2' until all the $-1$'s are moved once.

\vskip0.15cm
\noindent 
{\em Remark 4}.
Consider the $\gehn \neq A^{(1)}_n$ case and interpret the local states as follows:
\begin{center}
\begin{tabular}{c|c}
$B$ & Entry in the box \\\hline 
$1$ & empty \\[5pt]
$j \neq \pm 1$
& particle of color $j$ \\[5pt]
$-1$
&bound state of $j$ and $j^\ast$\\[5pt]
\end{tabular}
\end{center}
Since the set $B \setminus \{1, -1\}$ is invariant under the 
interchange $j \leftrightarrow j^\ast$,
the state $j^\ast$ may be viewed as the anti-particle of $j$ and vice versa.
In this sense the bound state $-1$ is neutral, and 
especially $0$ and $\emptyset$ represent the `self-neutral' particles
that can still form a bound state with another one.
Under this interpretation, the above algorithm for $K_j$ 
describes the motion of right-moving particles of color $j$ 
seeking the empty box or a free partner, 
{\it i.e.}, an anti-particle $j^\ast$ not yet paired with the other $j$'s, 
to form a neutral bound
state within a box.
\vskip0.2cm

The algorithm for $K_j$ can also be stated in terms of local rules,
which we shall now explain.
For $j \in J$ we introduce a map  $L_j: (\Z_{\ge 0}) \times B 
\rightarrow B \times (\Z_{\ge 0})$ as follows. 
Let \raise15pt\hbox{the diagram }
\batten{l}{b}{b'}{l'}{j}
\raise15pt\hbox{
denote $L_j:(l,b) \mapsto (b',l')$. 
($j$ is attached to the horizontal line.)}
\begin{enumerate}
\item
$j \notin \{1,0,-1,\emptyset\}$ case. Assume 
$l \in \mathbb{Z}_{\ge 0},\,b \in B
\backslash \{ j, -j, 1, -1 \}$. \\
\batten{l}{j}{1}{l\!+\!1}{j}\hspace{15pt}
\batten{l\!+\!1}{-j}{-1}{l}{j}\hspace{-3pt}
\batten{0}{-j}{-j}{0}{j}\hspace{-3pt}
\batten{l}{-1}{-j}{l\!+\!1}{j}\hspace{15pt}
\batten{l\!+\!1}{1}{j}{l}{j}\hspace{-3pt}
\batten{0}{1}{1}{0}{j}\hspace{-3pt}
\batten{l}{b}{b}{l}{j}
\item
$j \in \{0,\emptyset\}$ case. Assume 
$l \in \mathbb{Z}_{\ge 0},\,
b \in B \backslash \{ j, 1, -1 \}$. \\
\batten{l}{-1}{1}{l\!+\!2}{j}\hspace{15pt}
\batten{0}{j}{1}{1}{j}\hspace{15pt}
\batten{l\!+\!1}{j}{j}{l\!+\!1}{j}\hspace{25pt}
\batten{l\!+\!2}{1}{-1}{l}{j}\hspace{0pt}
\batten{1}{1}{j}{0}{j}\hspace{0pt}
\batten{0}{1}{1}{0}{j}\hspace{0pt}
\batten{l}{b}{b}{l}{j}\\
\item
$j=-1$ case. Assume 
$l \in \mathbb{Z}_{\ge 0},\,b \in B
\backslash \{ 1, j \}$. \\
\batten{l}{j}{1}{l\!+\!1}{j}\hspace{20pt}
\batten{l\!+\!1}{1}{j}{l}{j}\hspace{0pt}
\batten{0}{1}{1}{0}{j}\hspace{0pt}
\batten{l}{b}{b}{l}{j}\\
\end{enumerate}
For $\gehn = A^{(1)}_n$ we only have the third case with  
the $j$'s understood as $j \in \{2,3,\ldots, n+1 \}$.
See also \cite{HIK}.

Given an automaton state $(\ldots, b_i, b_{i+1}, \ldots) \in W$,
there exists an integer $m$ such that $b_{m'} = 1$ for all $m' < m$
owing to the boundary condition  (\ref{eq:W}).
Fix any such $m$. 
Then the operator $K_j : W \rightarrow W$ maps 
$(\ldots, b_i, b_{i+1}, \ldots)$ to $(\ldots, c_i, c_{i+1}, \ldots)$,
where $c_{m'} = b_{m'}$ for all $m' < m$.
The remaining $c_{m}, c_{m+1},\ldots$ are determined by the composition of 
$L_j$'s  as

\begin{center}
\begin{equation}\label{eq:LL}
\begin{picture}(80,20)(-12,0)
	\put(0,0){\vector(1,0){70}}
	\multiput(10,0)(20,0){3}{
		{\thicklines\put(0,10){\vector(0,-1){20}}}
		\put(0,4){\line(1,-1){4}}
		\put(2.5,2.5){\makebox(0,0)[lb]{{\tiny$j$}}}
	}
	\put(10,12){\makebox(0,0)[b]{$b_m$}}
	\put(30,12){\makebox(0,0)[b]{$b_{m+1}$}}
	\put(50,12){\makebox(0,0)[b]{$\cdots$}}
	\put(10,-12){\makebox(0,0)[t]{$c_m$}}
	\put(30,-12){\makebox(0,0)[t]{$c_{m+1}$}}
	\put(50,-12){\makebox(0,0)[t]{$\cdots$}}
	\put(-2,0){\makebox(0,0)[r]{0}}
\end{picture}
\end{equation}
\end{center}
\vskip25pt
It is easy to see that the result is independent of the choice of $m$.
The nonnegative integers $l$ on the horizontal line stand for the number of 
color $j$ particles on the carrier.
The diagrams for $L_j$ are viewed as the loading and unloading process 
of the color $j$ particles when the carrier proceeds to the right.
They match the algorithm stated before.
Perhaps the above formulation of $K_j$ using $L_j$ is easier to program.

\section{Examples}\label{sec:3}

Let us present examples of the factorized dynamics.
We suppress the trivial left tail in the time evolution and only depict the part
that corresponds to the composition of (\ref{eq:LL}) vertically.
\begin{example}
$\gehn=D^{(1)}_4$. 
$T = K_2K_3K_4K_{-4}K_{-3}K_{-2}$. \\
\setlength{\unitlength}{10pt}
\begin{tabular}{rp{1pt}cp{1pt}cp{1pt}cp{1pt}cp{1pt}cp{1pt}cp{1pt}cp{1pt}cp{1pt}cp{1pt}cp{1pt}}
&&   -3& &   -2& &    1& &    -2& &   2& &    3& &    1& &    1& &    1& \\
\yajirusiK{-2}&
0&\batu{-2}&0&\batu{-2}&1&\batu{-2}&0&\batu{-2}&1&
\batu{-2}&0&\batu{-2}&0&\batu{-2}&0&\batu{-2}&0&\batu{-2}&0\\
&&    -3& &   1& &  -2& &  1& & -1& &  3& &   1& &    1& &    1& \\
\yajirusiK{-3}&
0&\batu{-3}&1&\batu{-3}&0&\batu{-3}&0&\batu{-3}&0&
\batu{-3}&1&\batu{-3}&0&\batu{-3}&0&\batu{-3}&0&\batu{-3}&0\\
&&  1& &   -3& &  -2& & 1& &   3& &  -1& &  1& &    1& &    1& \\
\yajirusiK{-4}&
0&\batu{-4}&0&\batu{-4}&0&\batu{-4}&0&\batu{-4}&0&
\batu{-4}&0&\batu{-4}&1&\batu{-4}&0&\batu{-4}&0&\batu{-4}&0\\
&&  1& &  -3& &  -2& &   1& &  3& &  4& &  -4& &   1& &   1& \\
\yajirusiK{4}&
0&\batu{4}&0&\batu{4}&0&\batu{4}&0&\batu{4}&0&
\batu{4}&0&\batu{4}&1&\batu{4}&0&\batu{4}&0&\batu{4}&0\\
&&  1& &  -3& &  -2& &   1& &  3& &  1& &  -1& &   1& &   1& \\
\yajirusiK{3}&
0&\batu{3}&0&\batu{3}&0&\batu{3}&0&\batu{3}&0&
\batu{3}&1&\batu{3}&0&\batu{3}&1&\batu{3}&0&\batu{3}&0\\
&&  1& &  -3& &  -2& &   1& &  1& &  3& &  -3& &  3& &   1& \\
\yajirusiK{2}&
0&\batu{2}&0&\batu{2}&0&\batu{2}&0&\batu{2}&0&
\batu{2}&0&\batu{2}&0&\batu{2}&0&\batu{2}&0&\batu{2}&0\\
&&  1& &  -3& &  -2& &   1& &  1& &  3& &  -3& &  3& &   1& \\
\end{tabular}\\
\[
T: (-3,-2,1,-2,2,3,1,1,1,\ldots) \mapsto
(1,-3,-2,1,1,3,-3,3,1,\ldots).
\]
\end{example}

\vskip10pt

\begin{example}
$\gehn=B^{(1)}_3$. $T = K_2K_3K_0K_{-3}K_{-2}$.\\
\setlength{\unitlength}{10pt}
\begin{tabular}{rp{1pt}cp{1pt}cp{1pt}cp{1pt}cp{1pt}cp{1pt}cp{1pt}cp{1pt}cp{1pt}cp{1pt}cp{1pt}cp{1pt}cp{1pt}}
&&   -1& &   -3& &    0& &    3& &   -2& &    1& &    1& &    1& &    1& & 1&\\
\yajirusiK{-2}&
0&\batu{-2}&1&\batu{-2}&1&\batu{-2}&1&\batu{-2}&1&\batu{-2}&2&
\batu{-2}&1&\batu{-2}&0&\batu{-2}&0&\batu{-2}&0&\batu{-2}&0\\
&&    2& &   -3& &    0& &    3& &    1& &   -2& &   -2& &    1& &    1& & 1&\\
\yajirusiK{-3}&
0&\batu{-3}&0&\batu{-3}&1&\batu{-3}&1&\batu{-3}&0&\batu{-3}&0&
\batu{-3}&0&\batu{-3}&0&\batu{-3}&0&\batu{-3}&0&\batu{-3}&0\\
&&    2& &    1& &    0& &   -1& &    1& &   -2& &   -2& &    1& &    1& & 1&\\
\yajirusiK{0}&
0&\batu{0}&0&\batu{0}&0&\batu{0}&1&\batu{0}&3&\batu{0}&1&
\batu{0}&1&\batu{0}&1&\batu{0}&0&\batu{0}&0&\batu{0}&0\\
&&    2& &    1& &    1& &    1& &   -1& &   -2& &   -2& &    0& &    1& & 1&\\
\yajirusiK{3}&
0&\batu{3}&0&\batu{3}&0&\batu{3}&0&\batu{3}&0&\batu{3}&1&
\batu{3}&1&\batu{3}&1&\batu{3}&1&\batu{3}&0&\batu{3}&0\\
&&    2& &    1& &    1& &    1& &   -3& &   -2& &   -2& &    0& &    3& & 1&\\
\yajirusiK{2}&
0&\batu{2}&1&\batu{2}&0&\batu{2}&0&\batu{2}&0&\batu{2}&0&
\batu{2}&0&\batu{2}&0&\batu{2}&0&\batu{2}&0&\batu{2}&0\\
&&    1& &    2& &    1& &    1& &   -3& &   -2& &   -2& &    0& &    3& & 1&\\
\end{tabular}\\
\[
T: (-1,-3,0,3,-2,1,1,1,1,1,\ldots) \mapsto 
(1,2,1,1,-3,-2,-2,0,3,1,\ldots).
\]
\end{example}

\vskip20pt

\begin{example}\label{ex:A24}
$\gehn=A^{(2)}_4$. $T = K_2K_{-1}K_{-2}K_{\emptyset}$.\\
\setlength{\unitlength}{10pt}
\begin{tabular}{rp{1pt}cp{1pt}cp{1pt}cp{1pt}cp{1pt}cp{1pt}cp{1pt}cp{1pt}cp{1pt}cp{1pt}cp{1pt}cp{1pt}cp{1pt}}
&&   -1& &   -2& &$\emptyset$&&    2& &$\emptyset$&&   -2& &    1& &    1& &    1&&1& \\
\yajirusiK{\emptyset}&
0&\batu{\emptyset}&2&\batu{\emptyset}&2&\batu{\emptyset}&2&\batu{\emptyset}&2&\batu{\emptyset}&2&\batu{\emptyset}&2&\batu{\emptyset}&0&\batu{\emptyset}&0&\batu{\emptyset}&0&\batu{\emptyset}&0\\
&&    1& &   -2& &$\emptyset$&&    2& &$\emptyset$&&   -2& &   -1& &    1& &    1& &1&\\
\yajirusiK{-2}&
0&\batu{-2}&0&\batu{-2}&1&\batu{-2}&1&\batu{-2}&0&\batu{-2}&0&
\batu{-2}&1&\batu{-2}&2&\batu{-2}&1&\batu{-2}&0&\batu{-2}&0\\
&&    1& &    1& &$\emptyset$&&   -1& &$\emptyset$&&    1& &    2& &   -2& &   -2& &1&\\
\yajirusiK{-1}&
0&\batu{-1}&0&\batu{-1}&0&\batu{-1}&0&\batu{-1}&1&\batu{-1}&1&
\batu{-1}&0&\batu{-1}&0&\batu{-1}&0&\batu{-1}&0&\batu{-1}&0\\
&&    1& &    1& &$\emptyset$&&    1& &$\emptyset$&&   -1& &    2& &   -2& &   -2& &1&\\
\yajirusiK{2}&
0&\batu{2}&0&\batu{2}&0&\batu{2}&0&\batu{2}&0&\batu{2}&0&
\batu{2}&1&\batu{2}&2&\batu{2}&1&\batu{2}&0&\batu{2}&0\\
&&    1& &    1& &$\emptyset$&&    1& &$\emptyset$&&   -2& &    1& &   -1& &   -1& &1&\\
\end{tabular}\\
\[
T: 
(-1,-2,\emptyset,2,\emptyset,-2,1,1,1,1,\ldots) \mapsto
(1,1,\emptyset,1,\emptyset,-2,1,-1,-1,1,\ldots).
\]

\vskip0.2cm
\noindent
Along the Steps 1-4 in Section \ref{sec:2}, 
the action of $K_{\emptyset}$ in the above goes as follows.

\vskip0.3cm

\noindent
{\setlength{\unitlength}{\BallWidth}
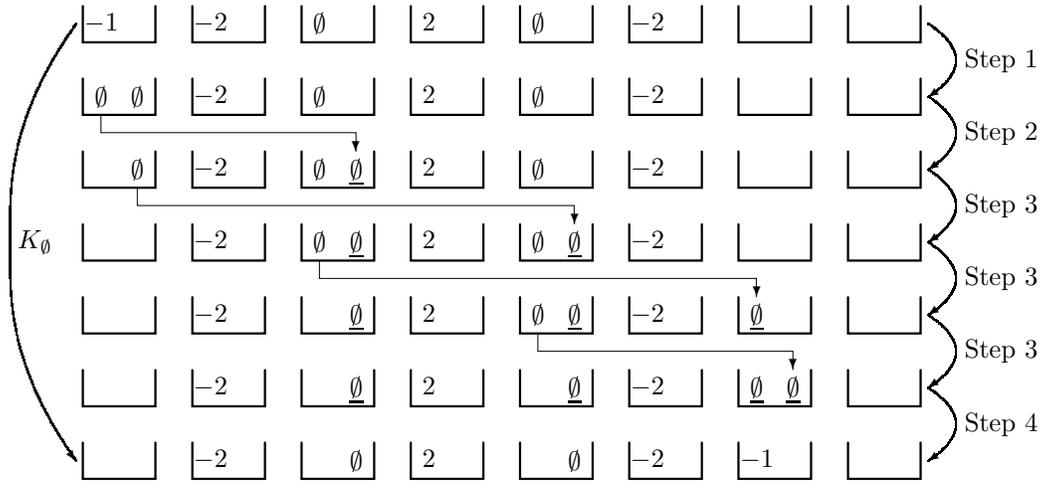
\begin{picture}(26,13)(-2,0)
\multiput(0,0)(0,2){7}{
	\multiput(0,0)(3,0){8}{\thicklines \path(0,1)(0,0)(2,0)(2,1)}
}
\qbezier(-0.2,12.5)(-2,10)(-2,7.5)
\put(-2,7.5){\line(0,-1){2}}
\put(-1.8,6.5){\makebox(0,0)[l]{$K_\emptyset$}}
\qbezier(-2,5.5)(-2,3)(-0.2,0.5)
\put(-0.2,0.5){\vector(2,-3){0}}
\put(0.5,12.8){\makebox(0,0)[t]{$-1$}}
\put(3.5,12.8){\makebox(0,0)[t]{$-2$}}
\put(6.5,12.8){\makebox(0,0)[t]{$\emptyset$}}
\put(9.5,12.8){\makebox(0,0)[t]{$2$}}
\put(12.5,12.8){\makebox(0,0)[t]{$\emptyset$}}
\put(15.5,12.8){\makebox(0,0)[t]{$-2$}}
\put(23,12.5){\yajirusiStep{1}}
\put(0.5,10.8){\makebox(0,0)[t]{$\emptyset$}}
\put(1.5,10.8){\makebox(0,0)[t]{$\emptyset$}}
\put(3.5,10.8){\makebox(0,0)[t]{$-2$}}
\put(6.5,10.8){\makebox(0,0)[t]{$\emptyset$}}
\put(9.5,10.8){\makebox(0,0)[t]{$2$}}
\put(12.5,10.8){\makebox(0,0)[t]{$\emptyset$}}
\put(15.5,10.8){\makebox(0,0)[t]{$-2$}}
\put(23,10.5){\yajirusiStep{2}}
\put(0.25,10){\ballmovingarrow{7}}
\put(1.5,8.8){\makebox(0,0)[t]{$\emptyset$}}
\put(3.5,8.8){\makebox(0,0)[t]{$-2$}}
\put(6.5,8.8){\makebox(0,0)[t]{$\emptyset$}}
\put(7.5,8.8){\makebox(0,0)[t]{$\underline{\emptyset}$}}
\put(9.5,8.8){\makebox(0,0)[t]{$2$}}
\put(12.5,8.8){\makebox(0,0)[t]{$\emptyset$}}
\put(15.5,8.8){\makebox(0,0)[t]{$-2$}}
\put(23,8.5){\yajirusiStep{3}}
\put(1.25,8){\ballmovingarrow{12}}
\put(3.5,6.8){\makebox(0,0)[t]{$-2$}}
\put(6.5,6.8){\makebox(0,0)[t]{$\emptyset$}}
\put(7.5,6.8){\makebox(0,0)[t]{$\underline{\emptyset}$}}
\put(9.5,6.8){\makebox(0,0)[t]{$2$}}
\put(12.5,6.8){\makebox(0,0)[t]{$\emptyset$}}
\put(13.5,6.8){\makebox(0,0)[t]{$\underline{\emptyset}$}}
\put(15.5,6.8){\makebox(0,0)[t]{$-2$}}
\put(23,6.5){\yajirusiStep{3}}
\put(3.5,4.8){\makebox(0,0)[t]{$-2$}}
\put(6.25,6){\ballmovingarrow{12}}
\put(7.5,4.8){\makebox(0,0)[t]{$\underline{\emptyset}$}}
\put(9.5,4.8){\makebox(0,0)[t]{$2$}}
\put(12.5,4.8){\makebox(0,0)[t]{$\emptyset$}}
\put(13.5,4.8){\makebox(0,0)[t]{$\underline{\emptyset}$}}
\put(15.5,4.8){\makebox(0,0)[t]{$-2$}}
\put(18.5,4.8){\makebox(0,0)[t]{$\underline{\emptyset}$}}
\put(23,4.5){\yajirusiStep{3}}
\put(3.5,2.8){\makebox(0,0)[t]{$-2$}}
\put(7.5,2.8){\makebox(0,0)[t]{$\underline{\emptyset}$}}
\put(9.5,2.8){\makebox(0,0)[t]{$2$}}
\put(12.25,4){\ballmovingarrow{7}}
\put(13.5,2.8){\makebox(0,0)[t]{$\underline{\emptyset}$}}
\put(15.5,2.8){\makebox(0,0)[t]{$-2$}}
\put(18.5,2.8){\makebox(0,0)[t]{$\underline{\emptyset}$}}
\put(19.5,2.8){\makebox(0,0)[t]{$\underline{\emptyset}$}}
\put(23,2.5){\yajirusiStep{4}}
\put(3.5,0.8){\makebox(0,0)[t]{$-2$}}
\put(7.5,0.8){\makebox(0,0)[t]{$\emptyset$}}
\put(9.5,0.8){\makebox(0,0)[t]{$2$}}
\put(13.5,0.8){\makebox(0,0)[t]{$\emptyset$}}
\put(15.5,0.8){\makebox(0,0)[t]{$-2$}}
\put(18.5,0.8){\makebox(0,0)[t]{$-1$}}
\end{picture}
}

\vskip0.2cm \noindent
As cautioned in {\em Remark} 2, one must distinguish
the moved $\emptyset$'s  and those not yet moved.
Here we have marked the moved ones with underlines.
\end{example}

\section{Solitons}\label{sec:4}
To save the space we shall write,  for example, 
$2^{y_2} (-3)^{y_{-3}}$ to signify the configuration of the 
local states 
$\overbrace{2,\ldots,2}^{y_2},\overbrace{-3,\ldots,-3}^{y_{-3}}$
in a segment of an automaton state 
for nonnegative integers $y_2$ and $y_{-3}$, {\it etc}.
In the sequel we assume $y_b \in \Z_{\ge 0}$ for any $b \in B$.
For each $\gehn$ consider the following configurations and define $v$ {}from them.

\noindent
$A^{(1)}_n: (n+1)^{y_{n+1}}\; \ldots \; 3^{y_3}\; 2^{y_2}$,\\
$\qquad \quad  v = \sum_{i=2}^{n+1}y_i$.
\vskip0.1cm

\noindent
$A^{(2)}_{2n-1}: (-2)^{y_{-2}}\; (-3)^{y_{-3}}\;\ldots \; 
(-n)^{y_{-n}}\; n^{y_n}\; \ldots \; 3^{y_3}\; 2^{y_2}$,\\
$\qquad \quad v = \sum_{i=2}^n(y_i+y_{-i})$.
\vskip0.1cm

\noindent
$A^{(2)}_{2n}: (\emptyset)^{y_{\emptyset}}
(-1)^{y_{-1}}\; (-2)^{y_{-2}}\;\ldots \; 
(-n)^{y_{-n}}\; n^{y_n}\; \ldots \; 2^{y_2}\; 1^{y_1},\quad
y_{\emptyset} \in \{0, 1\}, \; y_1 = y_{-1}$, \\
$\qquad \quad v = y_{\emptyset}+\sum_{i=1}^n(y_i+y_{-i})$.
\vskip0.1cm

\noindent
$B^{(1)}_{n}:  (-2)^{y_{-2}}\; (-3)^{y_{-3}}\;\ldots \; 
(-n)^{y_{-n}}\; 0^{y_0}\; n^{y_n}\; \ldots \; 3^{y_3}\; 2^{y_2}, 
\quad y_0 \in \{0, 1\}$,\\
$\qquad \quad v = y_0 + \sum_{i=2}^n(y_i+y_{-i})$.
\vskip0.1cm

\noindent
$C^{(1)}_{n}: 
(-1)^{y_{-1}}\; (-2)^{y_{-2}}\;\ldots \; 
(-n)^{y_{-n}}\; n^{y_n}\; \ldots \; 2^{y_2}\; 1^{y_1},\quad 
y_1 = y_{-1}$, \\
$\qquad \quad v = \sum_{i=1}^n(y_i+y_{-i})$.
\vskip0.1cm

\noindent
$D^{(1)}_n:  (-2)^{y_{-2}}\; (-3)^{y_{-3}}\;\ldots \; 
(-n)^{y_{-n}}\; n^{y_n}\; \ldots \; 3^{y_3}\; 2^{y_2},\quad 
y_n y_{-n} = 0$,\\
$\qquad \quad v = \sum_{i=2}^n(y_i+y_{-i})$.
\vskip0.1cm

\noindent
$D^{(2)}_{n+1}:(\emptyset)^{y_{\emptyset}}
(-1)^{y_{-1}}\; (-2)^{y_{-2}}\;\ldots \; 
(-n)^{y_{-n}}\; 0^{y_0}\; n^{y_n}\; \ldots \; 2^{y_2}\; 1^{y_1},\;
y_{\emptyset}, y_0 \in \{0, 1\}, \; y_1 = y_{-1}$, \\
$\qquad \quad v = y_{\emptyset}+y_0+\sum_{i=1}^n(y_i+y_{-i})$.
\vskip0.2cm
In an element of $W$, the above configuration is called 
a soliton with amplitude $v$ if it is 
surrounded by sufficiently many $1$'s $\in B$.
(For an amplitude $v$ soliton, $1^v$ in its right suffices.)
The data $\{y_b\}$ is the internal label of a soliton.
Under the time evolution $T$ (\ref{eq:T}), the amplitude $v$ solitons
propagate stably to the right with velocity $v$ 
if they stay sufficiently away {}from 
configurations different {}from  $1 \in B$ \cite{HKT1}.
It is a good exercise to check this claim by using the algorithm for $T$
in Section \ref{sec:2}.
Moreover, the following facts have been proved concerning 
the collisions of solitons with distinct amplitudes \cite{HKOTY}.
\begin{enumerate}
\item The set of amplitudes remains invariant under the collisions.
\item Two-soliton scattering rule is characterized by the 
combinatorial $R$ of $U_q({\mathfrak g}_{n-1})$. 
\item Multi-soliton scattering factorizes into the two-soliton ones.
\end{enumerate}
Leaving the precise statements to \cite{HKOTY}, we here 
include an example of collision of two solitons.

\begin{example} $\gehn = C^{(1)}_3$. 
To make the space uniform, we denote $-j$ by $\ol{j}$ for $j=1,2,3$.
Then the evolution of a two-soliton state under $T^t \,(0 \le t \le 5)$ 
is depicted as follows.\hfill
\begin{center}
t=0 : $\cdots 1  1 \bar{1} \bar{2} \bar{2} 3 1 1 1 1 1 \bar{2} \bar{3} 2 1 1 1 1 1 1 1 1 1 1 1 1 1 1 1 1 1 1 1 1 1 1 1 1\cdots $ \\
t=1 : $\cdots 1  1 1 1 1 1 1 \bar{1} \bar{2} \bar{2} 3 1 1 1 \bar{2} \bar{3} 2 1 1 1 1 1 1 1 1 1 1 1 1 1 1 1 1 1 1 1 1 1\cdots $ \\
t=2 : $\cdots 1  1 1 1 1 1 1 1 1 1 1 1 \bar{1} \bar{2} 3 1 1 \bar{2} \bar{2} \bar{3} 2 1 1 1 1 1 1 1 1 1 1 1 1 1 1 1 1 1\cdots $ \\
t=3 : $\cdots 1  1 1 1 1 1 1 1 1 1 1 1 1 1 1 1 \bar{1} 3 1 1 1 \bar{2} \bar{2} \bar{2} \bar{3} 2 1 1 1 1 1 1 1 1 1 1 1 1\cdots $ \\
t=4 : $\cdots 1  1 1 1 1 1 1 1 1 1 1 1 1 1 1 1 1 1 1 \bar{1} 3 1 1 1 1 1 \bar{2} \bar{2} \bar{2} \bar{3} 2 1 1 1 1 1 1 1\cdots $ \\
t=5 : $\cdots 1  1 1 1 1 1 1 1 1 1 1 1 1 1 1 1 1 1 1 1 1 1 \bar{1} 3 1 1 1 1 1 1 1 \bar{2} \bar{2} \bar{2} \bar{3} 2 1 1\cdots $
\end{center}
One observes that the initial two solitons $\bar{1} \bar{2} \bar{2} 3 1$ and 
$\bar{2} \bar{3} 2$ are scattered into the final solitons 
$\bar{1} 3 1$ and $\bar{2} \bar{2} \bar{2} \bar{3} 2$ with a phase shift.
In this way, the two-body scattering rule consists of 
the exchange of the internal labels of solitons and the phase shift.
\end{example}

\section{Relation to crystal theory}\label{sec:5}

The factorized dynamics in this paper is a translation of a
result in \cite{HKT2}.
By regarding the local states $-1,\ldots, -n$ here as $\ol{1}, \ldots, \ol{n}$,
$B$ can be identified with  the crystal $B_1$ as a set. 
In eq.(33) of \cite{HKT2}, take the integer $k$ as
$k = n-1$ for $D^{(1)}_n$ and $k=n$ for the other $\gehn$'s.
It leads to $a_k = 1$ (cf. Table I, II and eq.(11) therein)
for all the $\gehn$'s in question,
which implies the boundary condition $\ldots 1 1 1 \ldots$.
The time evolution in this case is given by 
$T = {\cal T}_{k+d}$, where the data $i_{k+d},\ldots, i_{k+1}$ 
in eq.(33) there reads
\begin{center}
\begin{tabular}{c|c}
$\gehn$ & $i_{k+d},\dots,i_{k+1}$ \\\hline 
$A^{(1)}_{n}$
&$2,3, \dots, n, n+1$\\[5pt]
$A^{(2)}_{2n-1}, B^{(1)}_n$
&$0,2,3,\dots,n-1,n,n-1,\dots,3, 2, 0$\\[5pt]
$A^{(2)}_{2n}, C^{(1)}_n, D^{(2)}_{n+1}$
&$1,2,\dots,n-1,n,n-1,\dots,2,1,0$\\[5pt]
$D^{(1)}_n$
&$0,2,3,\dots,n-2, n-1, n, n-2,n-3,\dots,3,2,0$\\[5pt]
\end{tabular}
\end{center}
\vskip1em
Under the convention ${\cal T}_k = id$, the operator 
$K_j$ in this paper has emerged {}from the formula
\begin{equation}\label{eq:relation}
K_{j_r} = {\cal T}_{k+r}({\cal T}_{k+r-1})^{-1}\quad 1 \le r \le d.
\end{equation}
We have verified that the right hand side yields 
the composition of $L_{j_r}$'s described in Section 
\ref{sec:2} by an explicit calculation. 
Similar factorized dynamics can be formulated 
corresponding to the other boundary conditions  
than $1 \in B$ in (\ref{eq:W}) by choosing 
a different $k$ in the above.
However such variations do not affect the qualitative feature 
of the automata.

\vspace{0.4cm}
\noindent
{\bf Acknowledgements} \hspace{0.1cm}
The authors thank K. Hikami, R. Inoue, M. Okado, T. Tokihiro 
and Y. Yamada for collaboration in their previous works.

\end{document}